\documentclass[AMA]{WileyNJD-v1}

\usepackage{color}

\usepackage{amssymb}
\usepackage{paralist}

\articletype{Article Type}%

\received{6 June 2017}
\revised{21 july 2017}
\accepted{18 August 2017}

\begin{document}

\title{Resource provisioning in Science Clouds: requirements and challenges}

\author[1]{{\'A}lvaro L{\'o}pez Garc{\'i}a}
\author[1,2]{Enol Fern{\'a}ndez-del-Castillo}
\author[1]{Pablo Orviz Fern{\'a}ndez}
\author[1]{Isabel Campos Plasencia}
\author[1]{Jes{\'u}s Marco de Lucas}

\authormark{A. L{\'o}pez Garc{\'i}a \textsc{et al}}

\address[1]{\orgdiv{Advanced Computing and e-Science Group}, \orgname{Instituto de F{\'i}sica de Cantabria (CSIC - UC)}, \orgaddress{\state{Santander}, \country{Spain}}}

\address[2]{\orgname{EGI Foundation}, \orgaddress{\state{Amsterdam}, \country{The Netherlands}}}

\corres{A. L{\'o}pez Garc{\'i}a, IFCA, Adva. los Castros s/n. 39005 Santander, Spain \email{aloga@ifca.unican.es}}

\abstract[Summary]{
Cloud computing has permeated into the IT industry in the last few years, and
it is nowadays emerging in scientific environments. Science
user communities are demanding a broad range of computing power to satisfy
high-performance applications needs, such as local clusters, High Performance
Computing {\color{blue}(HPC)} systems and computing grids.
{\color{blue}Different} workloads need from different computational models, and
the cloud is already considered as a promising paradigm.

The scheduling and allocation of resources is always a challenging matter in
any form of computation and clouds are not an exception. Science applications
have unique features that differentiate their workloads, hence their
requirements have to be taken into consideration to be fulfilled when building
a Science Cloud. This paper will discuss what are the main scheduling and
resource allocation challenges for any Infrastructure as a Service
{\color{blue}IaaS} provider supporting scientific applications.
}

\keywords{Scientific Computing, Cloud Computing, Science Clouds, Cloud Challenges}

\maketitle

\footnotetext{
    This is the pre-peer reviewed version of the following article:
    López García Á, Fernández-del-Castillo E, Orviz Fernández P, Campos
    Plasencia I, Marco de Lucas J. Resource provisioning in Science Clouds:
    Requirements and challenges. Softw Pract Exper. 2017;1-13,
    which has been published in final form at
    \url{https://doi.org/10.1002/spe.2544}. This article may be used for
    non-commercial purposes in accordance with Wiley Terms and Conditions for
    Self-Archiving.
}

\footnotetext{\textbf{Acknowledgments:} The authors want to acknowledge the
    support of the EGI-Engage (grant number 654142) and
    INDIGO-Data{\color{blue}cloud} (grant number 653549) projects, funded by
    the European Commission's Horizon 2020 Framework
    Programme{\color{blue}cloud}}

\section{Introduction}
\label{sec:introduction}

{\color{blue}Cloud computing} can be defined as ``\emph{a model for enabling ubiquitous,
convenient, on-demand network access to a shared pool of configurable computing
resources (e.g., networks, servers, storage, applications and services) that
can be rapidly provisioned and released with minimal management effort or
service provider interaction.}''~\cite{Mell2011}. This model allows many
enterprise applications to scale and adapt to the usage peaks without big
investments in hardware, following a \emph{pay-as-you-go} model without needing
an upfront commitment \cite{Armbrust2010} for acquiring new resources.

This computing paradigm has achieved great success in the IT industry but it is
still not common in the scientific computing field. Cloud computing leverages
virtualization \cite{Manvi2014} to deliver resources to the users, and the
associated performance degradation was traditionally considered as not
compatible with the computational science requirements \cite{Regola2010}.
However, nowadays it is widely accepted that virtualization introduces a CPU
overhead that can be neglected \cite{Barham2003,Ranadive2008,Campos2013}.
This
has been confirmed by several studies that have evaluated the performance of
the current cloud offerings both in public clouds such as Amazon EC2
\cite{Evangelinos2008,Walker2008,Afgan2010,Ostermann2010,Vockler2011,Rehr2011,Exposito2013,Oesterle2015}
or on private and community clouds
\cite{Campos2013,Rodriguez-Marrero2012,Hoffa2008,Gunarathne2011,gupta2011evaluation,Srirama2012,DeOliveira2012}.
{\color{blue}Moreover, other authors consider that the benefits that virtualization and
cloud computing introduces are often more important than a small performance
penalty \cite{Birkenheuer2012,Campos2013}.}

Therefore, and considering the expectations created around its promising
features, scientific communities are {\color{blue}starting to look with
interest in the cloud}. Some of the main characteristics are not novel ideas as they are
already present in current computing environments \cite{Foster2008}: academic
researchers have used shared clusters and supercomputers since long, and they
are being accounted for their usage in the same pay-per-use basis ---i.e.
without a fixed fee or upfront commitment--- based on their CPU-time and
storage consumption. Nevertheless, facts such as the customized environments,
resource abstraction and elasticity can fill some of the existing gaps in the
current scientific computing infrastructures \cite{Wang2008,Foster2008}.

Besides, current cloud middleware is designed to satisfy the industry needs. In a
commercial cloud provider users are charged in a pay-as-you-go basis, so the
customers pay according to their resource consumption. A commercial resource
provider might not worry about the \emph{actual} usage of the resources, as
long as they are getting paid by the consumed capacity, even if they are idle
resources. This situation is not acceptable in scientific facilities where the
maximum utilization of the resources is an objective. Idle resources are an
undesirable scenario if it prevents other users from accessing and using the
infrastructure. Access to scientific datacenters is not based on a pay per use
basis, as user communities are granted with an average capacity over long
periods of time. This capacity, even if accounted, is not paid by the users,
but it is rather supported by means of long-term grants or agreements.

In traditional scientific datacenters users execute their tasks by means of
the well known batch systems, where the jobs are normally time-bounded (i.e. they
have a specific duration). Different policies are then applied to adjust the
job priorities so that the resources are properly shared between the different
users and groups. Even if the user does not specify a duration, a batch system
is able to stop its execution after a given amount of time, configured by the
resource provider.

However, there is no such \emph{duration} concept in the cloud model, where a
virtual machine is supposed to live as long as the user wants. Users may not
stop their instances when they have finished their job (they are not getting
charged for them), ignoring the fact that they are consuming resources that may
be used by other groups. Therefore, resource providers have to statically
partition their resources so as to ensure that all users are getting their
share in the worst situation. This leads to an underutilization of the
infrastructure, since a usage spike from a group cannot be satisfied by idle
resources assigned to another group.

\begin{figure}[h!]
    \centering
    \includegraphics[width=0.6\linewidth]{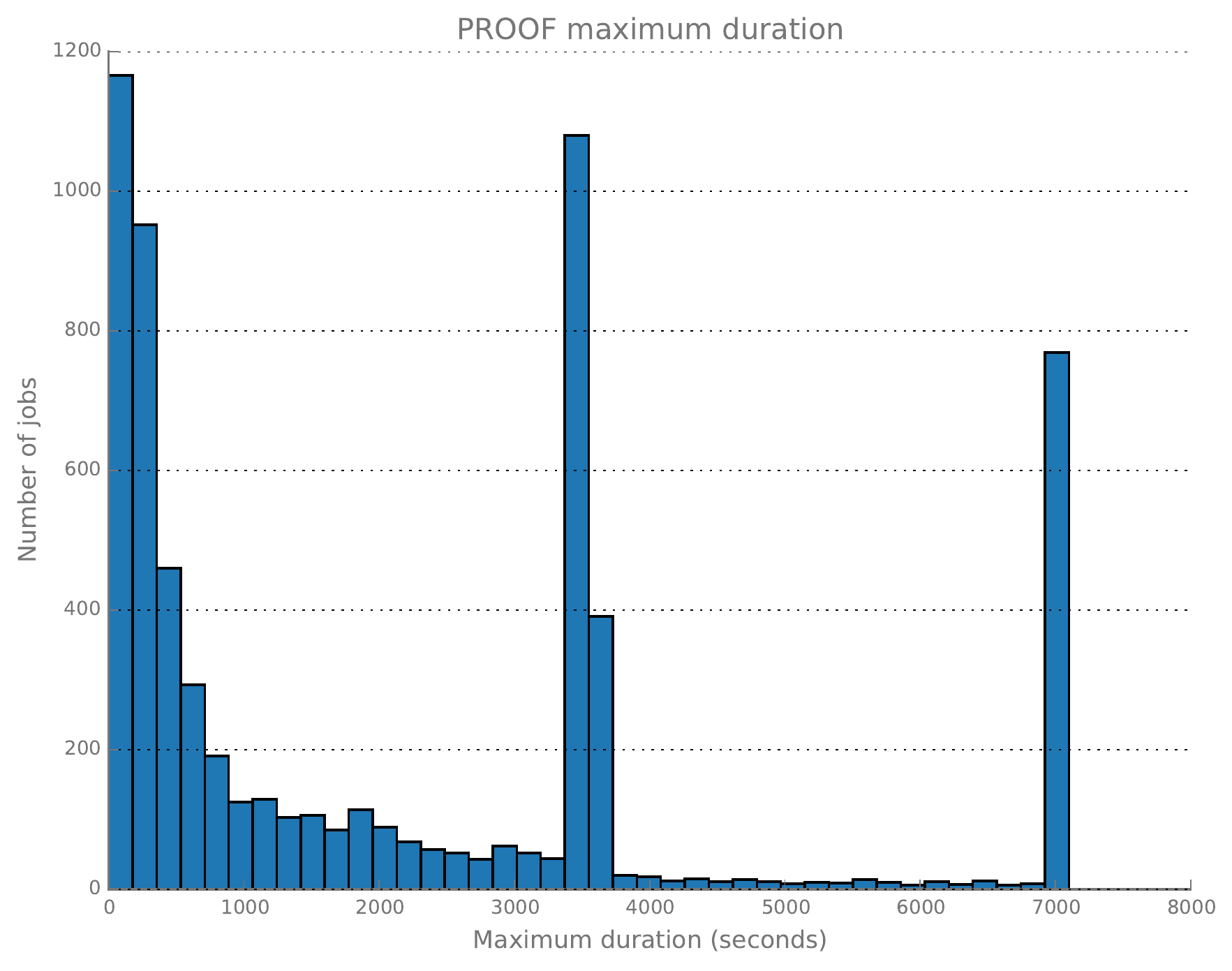}
    \caption{PROOF task duration.}
    \label{fig:proof duration}
\end{figure}

\begin{figure}[h!]
    \centering
    \includegraphics[width=0.6\linewidth]{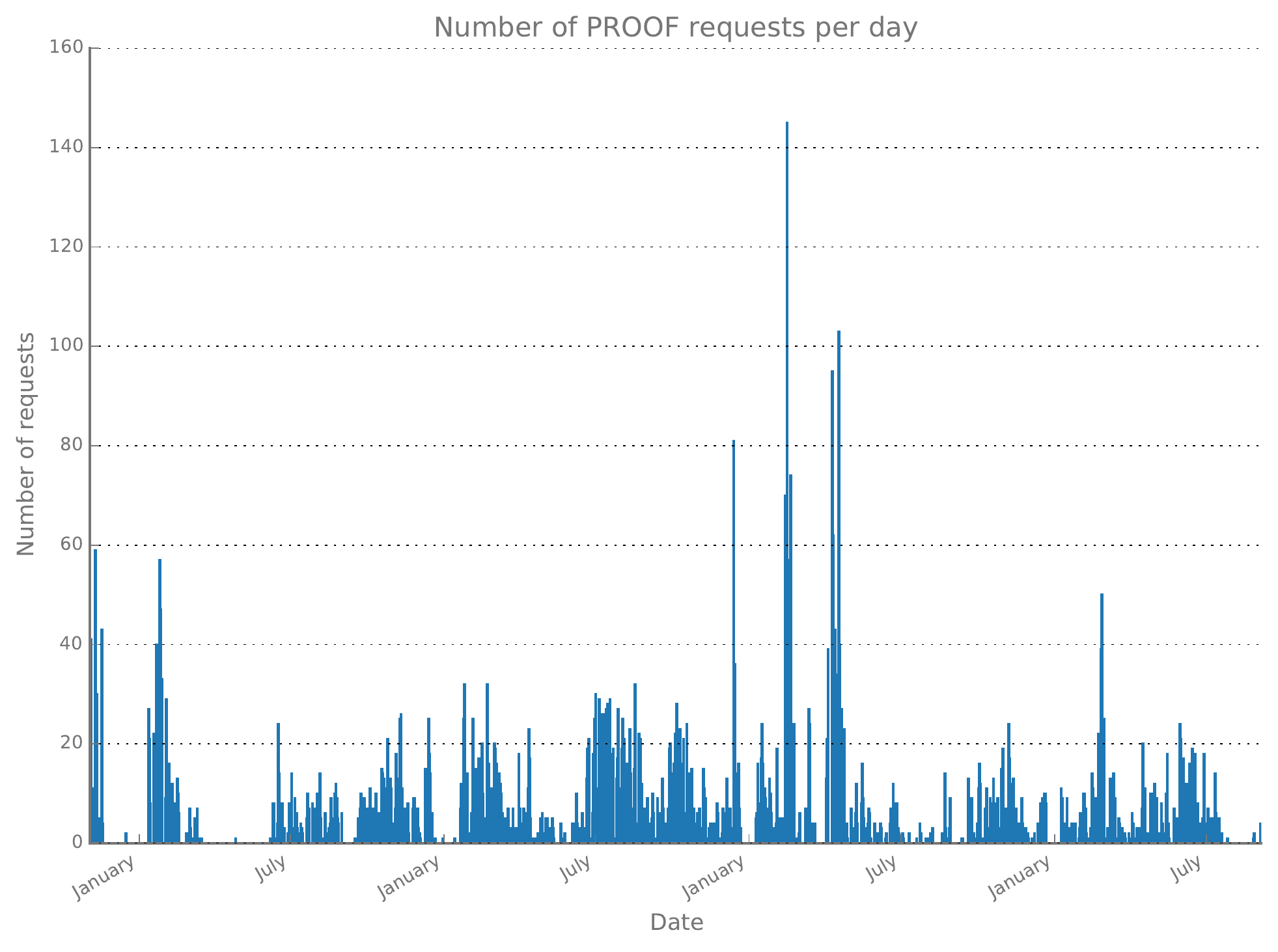}
    \caption{PROOF daily request pattern for a three and a half year period.}
    \label{fig:proof reqs}
\end{figure}

{\color{blue}To illustrate this problem} we have collected the usage patterns for several
months from a batch system specially configured to support this kind of tasks,
regarding one application widely used by the High Energy Physics (HEP)
community: the Parallel ROOT Facility (PROOF) \cite{Antcheva2009}. This tool is
used to perform interactive analysis of large datasets produced by the current
HEP experiments.  Figure~\ref{fig:proof duration} shows the number of requests
regarding the task duration. As it can be seen, all the requests can be
considered short-lived, since its maximum duration is below 2 hours, with the
highest concentration being below 1 hour. Figure~\ref{fig:proof reqs} depicts
the request pattern for a $3.5$ year period. As it can be seen, this kind of
jobs are executed in bursts or waves, meaning that a set of users will have a
high demand of resources for short periods of time ---i.e. when an analysis is
at a final stage.

This kind of usage (i.e. short lived executions that are not constant over the
time) is quite common for scientific applications
\cite{Campos2013,nilsson2014extending,albor2015multivariate,Voorsluys2011,Juve2008}
and presents a demanding challenge for resource providers. It is
needed to deliver enough computing capacity for absorbing  this kind or
request, minimizing the reserved resources that will be idle for long periods
of time.

Implementing an effective scheduling and resource allocation policies to ensure
that the elasticity is perceived as true is a challenging task. The issue in
this context is the need of maximizing the utilization of the Infrastructure
as a Service (IaaS) resources, so that a minimal amount of physical resources
are provisioned and maintained. An allocation policy that is driven by a
resource provider decision can result in a low value from the user standpoint,
whereas an allocation under user control may result in a high cost for the
provider \cite{Manvi2014}.

In addition, satisfying elastic requests in an efficient way is not the sole
challenge that a resource provider will face. Scientific applications have
unique requirements, therefore {\color{blue}Science Clouds} shall provide unique features
and face unique challenges. In this work we will focus on a gap analysis for a
scientific IaaS provider, so that an effective resource allocation can be done.
We will not focus in the mere virtual to physical resource mapping, but we will
also cover other resource allocation {\color{blue}problematic}.

{\color{blue}
The rest of the paper is organized as follows. In Section~\ref{sec:related} we
will perform a review of the related work in the area. In
Section~\ref{sec:challenges} we will cover the open challenges that we have
identified from the resource provisioning point of view. Finally, our
conclusions are presented in Section~\ref{sec:conclusions}.
}

\section{Related work}
\label{sec:related}

To the best of our knowledge, there are not so many studies considering the
resource allocation {\color{blue}problematic} from the resource provider point
of view that take into account the specificity of the scientific application
requirements.

There is a considerable amount of research works addressing cloud resource
provisioning and scheduling from the user or consumer perspective
\cite{Chaisiri2011,Voorsluys2011,Voorsluys2012,Huang2013}. Some authors
have studied how to implement hybrid provisioning of resources between several
{\color{blue}cloud} providers \cite{Sotomayor2009,Montero2011}, or even between different
computing infrastructures such as grids and clouds \cite{Calheiros2012}. The
workflow model is widely used in many scientific computing areas, and there is
a vast amount of studies regarding the feasibility and challenges of executing
workflows in the cloud
\cite{Hardt2012,Lee2015,Smanchat2015,Rodriguez2014,Lin2013,jung2014workflow,Szabo2014}.

{\color{blue}
The systematic survey performed by Chauhan et al. \cite{Chauhan2017} identified
some challenges for High Performance Computing (HPC) and scientific computing
in the cloud. More specifically, the survey points to the work by Somasundaram
and Govindarajan \cite{Somasundaram2014} where the authors develop a framework
focused on the execution of HPC applications in cloud environments by managing
cloud resources where the user application is dispatched.}

Regarding resource provisioning strategies from the provider standpoint,
Sotomayor et al. studied how to account and manage the overheads introduced the
virtual resources management \cite{Sotomayor2006}. Hu et al. \cite{Hu2009}
studied how to deliver a service according to several agreed Service Level
Agreements (SLAs) by using the smallest number of resources. Garg et al.
\cite{Garg2011} presented how to deal with SLAs that imply interactive and
non-interactive applications. Cardonha et al. \cite{Cardonha2013} proposed a
patience-aware scheduling that take into account the user's level of tolerance
(i.e. the patience) to define how to deliver the resources to the users.

There is {\color{blue} large number of research works} regarding energy aware
resource provisioning in the clouds
\cite{Buyya2010,Orgerie2011,Beloglazov2012}. Smith et al.  \cite{Smith2011}
modelled how different workloads affected energy consumption, so that an
accurate proper power prediction could be made to perform an efficient
scheduling. Several authors have studied how the consolidation of virtual
servers in a cloud provider could lead to a reduction of the energy consumption
\cite{Corradi2014,Srikantaiah2008}. This fact can be used to increase the
revenues by implementing energy-aware resource allocation policies
\cite{Mazzucco2011}.

{\color{blue}
Kune et al. \cite{Kune2016} elaborated an exhaustive taxonomy of big data
computing, including a discussion on the existing challenges and approaches for
big data scheduling ({\color{blue}among} others). This work also includes an study of the
underpinning technologies for big data cloud computing, as long as a gap
analysis in the current architectures and systems.}

{\color{blue}Manvi et al.} \cite{Manvi2014} performed an exhaustive review of the resource
provisioning, allocation and mapping problems for a IaaS resource provider,
stating some open challenges like
\begin{inparaenum}[\bfseries i)]
\item how to design a provisioning algorithm for optimal resource utilization
    based on arrival data;
\item how and when to reallocate VMs;
\item how to minimize the cost of mapping the request into the underlying
    resources;
\item how to develop models that are able to predict applications performance
\end{inparaenum}; among many others.

{\color{blue}On the other hand, there are previous studies} regarding the general
challenges for Science Clouds. The work by Blanquer et al.~\cite{Blanquer}, in
the scope of the VENUS-C project, evaluated the requirements of scientific
applications by performing a broad survey of scientific applications within the
project. Their study showed that the {\color{blue}cloud computing model} is perceived as
beneficial by the users (being one of the key expectations the elasticity),
although some drawbacks need to be tackled so as to improve its adoption (such
as interoperability, learning curve, etc.).

Juve et al. \cite{Juve2010} outlines what is expected from a Science Cloud in
contrast with a commercial provider (shared memory, parallel applications,
shared filesytems) so as to effectively support scientific workflows. Besides,
it concluded that cloud can be beneficial for scientific users, assuming that
Science Clouds will be build ad-hoc for its users, clearly differing from
commercial offers.

{\color{blue}
The United States Department of Energy (DOE) Magellan project elaborated an
extensive report on the usage of cloud computing for science \cite{magellan} by
deploying several cloud infrastructures that were provided to some selected
mid-range computing and data intensive scientific applications. Their key
findings include
\begin{inparaenum}[\bfseries i)]
\item the identification of advantages of the cloud computing model, like
    the availability of customized environments for the user or flexible
    resource management;
\item the requirement of additional programming and system administration
    skills in order to adopt the cloud computing model;
\item the economic benefit of the cloud computing model from the provider
    perspective due to the resource consolidation, economies of scale and
    operational efficiency; and
\item some significant gaps that exist in several areas,
    including resource management, data, cyber-security and others.
\end{inparaenum}

Regarding this last finding, the study concluded that there are several open
challenges that science clouds need to address in order to ensure that
scientists can harness all the capabilities and potential that the cloud is
able to offer. These needs derive from the special requirements that scientific
applications have, and were collected in a further publication  by Ramakrishnan
et al.  \cite{Ramakrishnan2011}. The authors conclude that science clouds
\begin{inparaenum}[\bfseries i)]
    \item need access to low-latency interconnects and filesystems;
    \item need access to legacy data-sets;
    \item need MapReduce implementations that account for characteristics of
        science data and applications;
    \item need access to bare metal provisioning;
    \item need pre-installed, pre-tuned application software stacks;
    \item need customizations for site-specific policies; and
    \item need more sophisticated scheduling methods and policies.
\end{inparaenum}
Some of those findings are coincident with the gaps that we have identified in
this work specially those regarding with resource management (like access to
specialized hardware) and scheduling policies, but further elaboration is
needed on them.
}

\section{Resource provisioning in Science Clouds}
\label{sec:challenges}

Scientific workloads involve satisfying strong requirements. Resource
allocation for scientific applications appears then as a demanding task that
should take into consideration a number of hardware and software variables. As
scientific applications started to move to cloud solutions, this number of
requirements got increased: on top of the already existing needs, new
requirements arose from the defining characteristics that the new paradigm of
cloud computing offered to users: \textit{on-demand self-service provisioning}
needs richer computing capabilities definitions for applications with e.g.
very specific demanding hardware requirements like guaranteeing a minimum
network bandwidth for remote data access, commonly found in scientific
environments. In the same line, \textit{elastic provisioning} is required to be
highly customizable for the sake of minimizing customers' budgets and
administrative costs for the service providers. Granular and customizable
environments increase \emph{predictability} so that the providers can offer
performance guarantees to customers while estimating accurately the costs of
resource utilization. Elasticity needs to be rapid as well: reducing e.g.
\emph{instance startup} will benefit a fast (auto-)scaling of resources.

In this section we will also cover other non-cloud inherent scientific
requirements, most of which were traditionally tackled in previously proposed
computing paradigms, such as {\color{blue}grid} computing and HPC clusters. Science Clouds
will need to provide resource provisioning methods and policies to satisfy
complex requirements such as \emph{resource co-allocation} or
\emph{performance and data aware-based provisioning}. But popular open source
cloud frameworks do not count with schedulers that have built-in mechanisms and
policy-definition to satisfy these requirements. Cloud schedulers surely are
not meant to offer the advanced set of scheduling possibilities that a
standard batch system has, but they definitely need to address those
requirements commonly found in scientific computations. A clear example is the
execution of non-interactive applications. Batch executions are needed in
multiple scientific use cases, so it appears to be reasonable to add
\emph{\color{blue}flexible allocation policies} to deal with this type of executions.

In the following lines we elaborate on the above identified requirements and,
for some cases, depict what resource allocation challenges and solutions can be
applied within Science Clouds.


\subsection{Instance co-allocation}
\label{sec:challenges:co allocation}

Compute and data intensive scientific workloads tend to use parallel techniques
to improve their performance. Parallel executions are complex since they
require intercommunication between processes, usually located in distributed
systems, scenario in which resource provisioning task becomes even more
challenging. Based on the assumption that a provider is capable of satisfying a
request involving different instances, one have to consider the fact of
managing them as a whole so to assure that these instances are actually being
provisioned at the same time i.e. they are being \emph{co-allocated}
{\color{blue}(in this context,
instance co-allocation is not related with executing several instances in the
same physical node, but rather that the instances are provisioned to the user
at the same time)}. Proper
co-allocation policies should take into account network requirements,
such as satisfying low latencies and appropriate bandwidths, and
{\color{blue}fulfil} any
constraints imposed by the parallel framework being used, as e.g. OpenMPI's
intra-subnet allocation check \cite{Evangelinos2008}.

{\color{blue}
If a proper co-allocation mechanism is not in place, users and resource
providers would need to coordinate in order to pre-provision the required
instances \cite{Ismail2012}, therefore hindering the on demand and self-service experience
that is expected from a cloud system.}

In homogeneous and static environments, guaranteeing ordered co-allocation of
resources can be easily tackled, if compared to heterogeneous scenarios.
In the specific case of cloud computing, the flexibility that it introduces,
makes multi-resource allocation a challenging task that must take into
consideration not only the synchronized startup (see more at
Section~\ref{sec:challenges:startup}) of master and worker instances, but
also how these resources are geographically distributed and what are the
hardware constraints (network, cpu, memory) to be considered. Only by doing
this, parallel tasks provisioned in clouds would have a similar application
performance than what can be obtained with homogeneous ad-hoc resources, but
getting rid of the rigidity that the introduce.

\subsubsection{Instance co-allocation open challenges}
The open challenges in this area are as follows:
\begin{itemize}
    \item How to offer a proper SLA to ensure that instances need to be
        co-allocated.
    \item How to ensure that instances that need co-allocation are actually
        started at the same time.
    \item How to account (or not account) for instances that requiring
        co-allocation have been provisioned with an unacceptable delay. When a
        user is requiring this feature but the requirement cannot be fulfilled
        this should be taken into account.
    \item How to ensure that when the instances are already scheduled
        they are allocated within a time-frame. VM management introduces
        overheads and delays that should be taken into account to ensure a
        proper co-allocation.
\end{itemize}

\subsection{Licensed software management}

One of the major barriers scientists find when moving their applications to the
cloud relies in licensing troubles. Software vendors that count with policies
about how to deal with licensing in virtualized environments propose the usage
of Floating Network Licenses (FNL). These special licenses usually increment
costs, as they can be used by different virtual instances, and require the
deployment of license managers in order to be able to use the software in the
cloud infrastructures. Additionally, the license managers might need to be
hosted within a organization's network.

Using FNLs are the most popular solution provided by vendors, but the imposed
requirements mentioned above can be difficult to satisfy in some cases: hosting
a license manager is not always possible by some scientific communities and it
introduces maintenance costs, whose avoidance is one of the clear benefits of
moving to a cloud solution.

The need for a more straightforward way of getting licensed or proprietary
software to work in virtualized environments is a must that software vendors
should consider. In commercial cloud infrastructures, like Amazon AWS,
customers can make use of pre-configured images, license-granted, with the
proprietary software locally available and ready to use. At the time of writing,
Amazon AWS does not have agreements with all of the major software vendors, but
it appears as a neat and smooth solution that requires no extra work from the
end users side.

Besides the above administrative difficulties, the actual technical challenge
in resource allocation for licensed software is that cloud schedulers are not
license-aware. This gap needs to be filled by the cloud middleware stacks, as
it was solved years ago in HPC clusters.

\subsubsection{Licensed software management open challenges}
The open challenges in this area are as follows:
\begin{itemize}
    \item Persuade commercial vendors to release more flexible licensing
        methods, specific for the cloud.
    \item How to deal with license slots within the scheduler.
\end{itemize}

\subsection{Performance aware placement}

In order to improve resource utilization, cloud schedulers can be configured to
follow a fill-up strategy that might end up in multiple virtual machines
running concurrently on the same physical server. This scenario leads to
resource competition which surely will affect application performance. In this
regard, the scheduler {\color{blue}that is in charge of provisioning the
resources in Science Clouds needs to be performance-aware (or even
degradation-aware), so that performance demanding instances do not share the
physical resources with other instances that may impact its performance.
}

Several approaches have been raised in order to diminish degradation. Some do
not act directly on pro-active scheduling but instead in reactive reallocation
of the affected virtual instances by using underneath hypervisor capabilities
like live migration. But, instead of relying in monitoring the application
performance and take reallocation decisions based upon its degradation, a more
pro-active scheduling is needed so to improve the suitability of the resource
selection. Feeding the scheduler with more fine-grained hardware requirements,
provided by the user request, such as low-latency interconnects (e.g.
Infiniband, 10GbE) or GPGPU {\color{blue}\cite{Chen2017}} selection, provides a
better resource categorization and, consequently, will directly contribute to a
more efficient execution of the application. To accomplish this, the
specialized hardware must be exposed into the virtual instances, by means of
PCI passthrough with IOMMU or Single Root I/O Virtualization (SR-IOV)
techniques, and eventually managed by the cloud middleware using the underlying
virtualization stack.

{\color{blue}
Therefore, consolidating virtual machines into the same physical host should
not be applied when the instances are executing performance demanding
applications.  Virtualization in these cases is used only a as a way to provide
customized environments for scientists. However, it should be noted that
science clouds can apply consolidation techniques for non demanding
applications, such as web portals or science gateways.}

\begin{figure}[h!]
    \centering
    \includegraphics[width=0.6\linewidth]{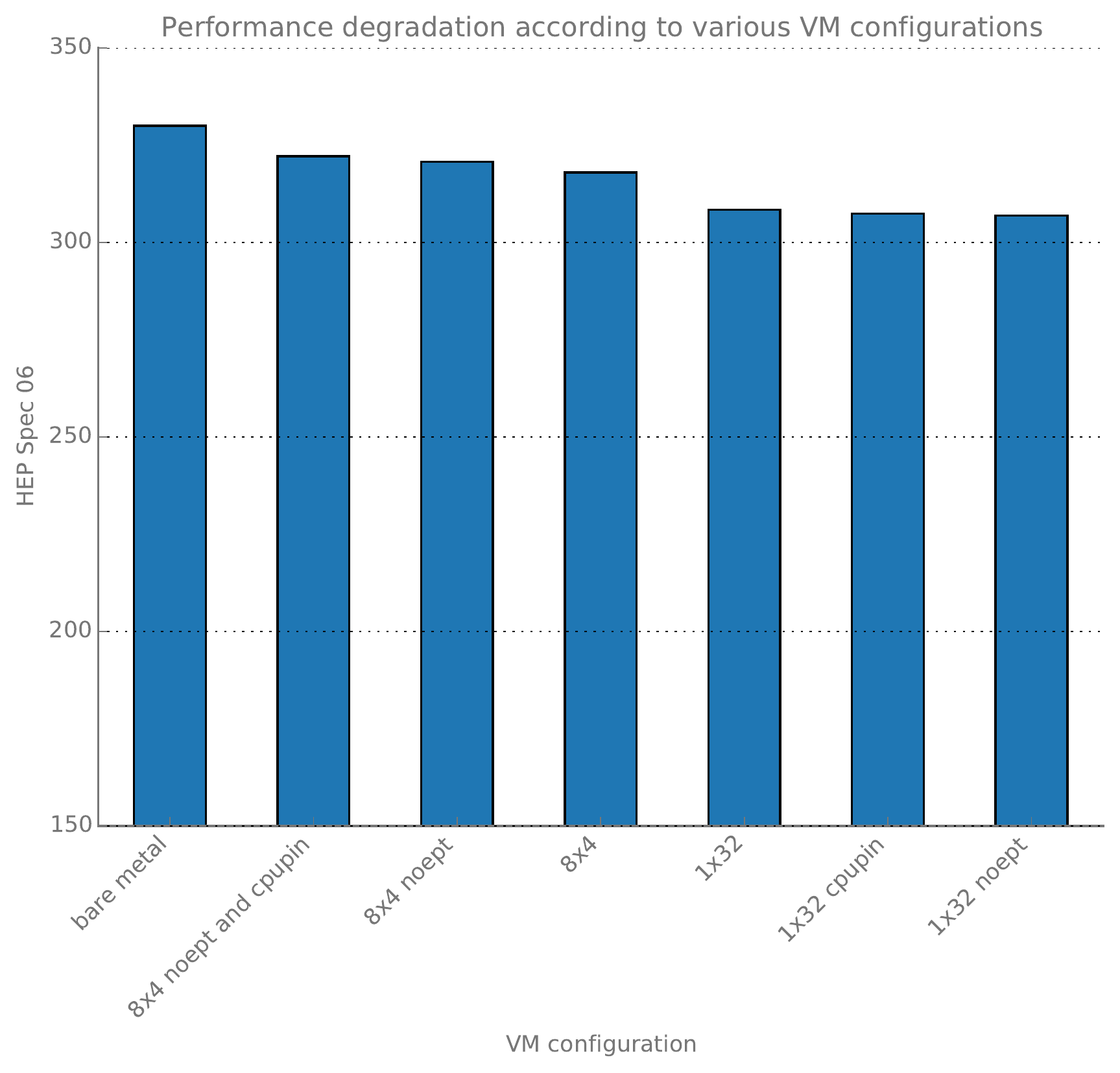}
    \caption{Aggregated performance regarding the
        HEP Spec 06 \cite{Michelotto2010} benchmark, taking into account
        different virtual machine sizes and configurations for one host. The
        physical node consists on a node with two 8-core Intel\circledR
        Xeon\circledR E5-2670 2.60GHz processors, 128GB RAM and the virtual
        machines were dimensioned so as to consume ---in aggregate--- all the
        resources available on the host. The label "noept" means that the Extended
        Page Tables (EPT) support has been disabled. The label "cpupin" means that the
        virtual CPUs have been pinned to the physical CPUs}
    \label{fig:performance degradation}
\end{figure}

The hypervisor providing the virtualization appears as an important factor when
measuring performance. It is widely accepted that virtualization introduces a
penalty when compared with {\color{blue}bare metal} executions. However this penalty depend on
how the hypervisor is being used. Figure~\ref{fig:performance degradation}
shows the degradation of the aggregated performance delivered by a physical
machine, using different vCPUs sizes.

{\color{blue}Science Clouds} need to deliver the maximum performance possible. Therefore, the
cloud middleware should take this fact into account, by implementing scheduling
policies that would help to prevent the above identified performance drops.

\subsubsection{Performance aware placement open challenges}
The open challenges in this area are as follows:
\begin{itemize}
    \item How to minimize the performance loss when scheduling various virtual
        machines inside one host.
    \item How to proactively scheduling could be used to minimize  resource
        competition.
    \item How to detect performance interferences between VMs and
        take {\color{blue}appropriate} actions (like live migration) to minimize them.
    \item How to redistribute the running instances between the resources
        without impacting the running applications.
    \item \color{blue} How to apply consolidation techniques that do not
        interfere with a scheduling strategy ensuring that performance demanding
        applications are not executed in a time sharing manner.
\end{itemize}

\subsection{Data-aware scheduling}

Several scientific disciplines ---such as High Energy Physics (HEP), Astronomy
or Genomics just to cite some of them--- generate considerably large amounts of
data (in the order of Petabytes) that need to be analyzed. Location and access
modes have clear impacts to data-intensive applications
{\color{blue}\cite{Shamsi2013,Kosar2006,Tan2013,Kune2016}}
and any platform that supports these kind applications should provide
data-locality and data-aware scheduling to reduce any possible bottlenecks that
may even prevent the actual execution of the application.

Storage in clouds is normally decoupled from the virtual machines and attached
during runtime upon user's demand. This poses a bigger challenge to the
scheduler since the location of data to be accessed is not known a priori by
the system. {\color{blue}Science Clouds} should be able to provide
high-{\color{blue}bandwidth} access to
the data, which is usually accessed over the network (e.g. block storage may use
ATA over Ethernet or iSCSI; object storage usually employs HTTP). This may
require enabling the access to specialized hardware from the virtual machines
(e.g. Infiniband network) or re-locating the virtual machines to hosts with
better connectivity to the data sources. Data-locality can also be improved by
using caches at the physical nodes that host the VMs, by replicating locally
popular data hosted externally to the cloud provider, or by leveraging tools
like CernVMFS \cite{Blomer2012} that deliver fast access to data using HTTP
proxies.

\subsubsection{Data-aware scheduling open challenges}
The open challenges in this area are as follows:
\begin{itemize}
    \item How to take into account cloud data management
        {\color{blue}specificities} when
        scheduling machines.
    \item How to ensure that the access delivers high performance for the
        application being executed.
\end{itemize}

\subsection{Flexible resource allocation policies}

Long-running tasks are common in computational science. Those kind of workloads
do not require from interactivity and normally are not time-bounded. Such tasks
can be used as opportunistic jobs that fill the computing infrastructure usage
gaps, leading to a better utilization of resources.

In traditional scientific
datacenters and time-sharing facilities this is normally done in by means of
several techniques, such as backfilling, priority adjustments, task preemption
and checkpointing. Some of these techniques require that the tasks are
time-bounded, but in the cloud a virtual machine will be executed as long as
the user wants.

Commercial cloud providers have tackled this issue implementing the so called
spot instances or {\color{blue}preemptible} instances. This kind of instances can be
terminated without further advise by the provider if some policy is violated
(for example, if the resource provider cannot satisfy a normal request ---in
the {\color{blue}preemptible} case--- or because the user is paying a prize that is
considered too low over a published price baseline ---in the spot mode, where
the price is governed by a stock-options like market.

The usage of this kind of instances in Science Clouds could make possible that
the infrastructure is filled with opportunistic \cite{Hategan2011} jobs that
can be stopped by higher priority tasks, such as interactive demands. The
Vacuum computing model \cite{McNab2014}, where resources appear
in the vacuum to process some tasks and then disappear is an ideal candidate to
leverage this kind of spot instances. Tools such as Vcycle \cite{web:vcycle}
{\color{blue} or SpotOn \cite{Subramanya:2015:SBC:2806777.2806851}
are already being used to profit from opportunistic usage in existing
commercial or  scientific infrastructures.}

\subsubsection{Flexible resource allocation policies open challenges}
The open challenges in this area are as follows:
\begin{itemize}
    \item How to maximize the resource utilization without preventing
        interactive users from accessing the infrastructure.
    \item How to specify dependencies between virtual machines so that
        workflows can be scheduled in a more easy way.
    \item How to account for resources that are suitable for being stopped or
        preempted.
    \item How to select the best instances that can be stopped to leave room
        for higer priority requests, with the compromise of reducing the
        revenue loss and with the smallest impact to the users.
\end{itemize}

\subsection{Performance predictability}

Popular open-source Infrastructure as a Service frameworks do not currently
expose mechanisms for customers to define a specific set of hardware
requirements that would guarantee a minimum performance when running their
applications in the cloud. Real time demanding or latency sensitive
applications are indeed seriously hit by this limitation, which appears as a
big obstacle for integrating this type of applications into clouds.

Computing capabilities provide only a magnitude of multi-threading efficiency
based on the number of virtual CPUs (vCPUs) selected. Customers are then tied
to a generic vCPU selection that may be mapped to different processors by the
underlying framework, in which case different performance results could be
obtained based on the same set of requirements. This unpredictability will be
increased whenever resource overcommit is in place, that could lead to CPU
cycle sharing among different applications.

Lack of network performance guarantees contribute also to unexpected
application {\color{blue}behaviour}. Enforcing network {\color{blue}Quality of
Service (QoS)}
to achieve customer-required network bandwidth can greatly improve application
predictability, but network requirement selection are seldom offered by cloud
providers \cite{Mogul2012}.

Improved performance predictability is a key requirement for users
\cite{Fakhfakh2014} but also to providers. The lack of predictability leads to
uncertainty \cite{Tchernykh2015}, a fact that should be mitigated for both
users and providers. An accurate provision of customer needs in
terms of computing and network capabilities will not only boost customer
experience but also will provide a clear estimation of cost based on the
different service levels that the resource provider can offer.

\subsubsection{Performance predictability open challenges}
The open challenges in this area are as follows:
\begin{itemize}
    \item How to expose enough granularity in the request specification without
        exposing the underlying {\color{blue}abstracted} resources.
    \item How to guarantee the performance predictability between different
        requests with the same hardware requirements.
\end{itemize}

\subsection{Short startup overhead}
\label{sec:challenges:startup}

\begin{figure}[h!]
    \centering
    \includegraphics[width=0.6\linewidth]{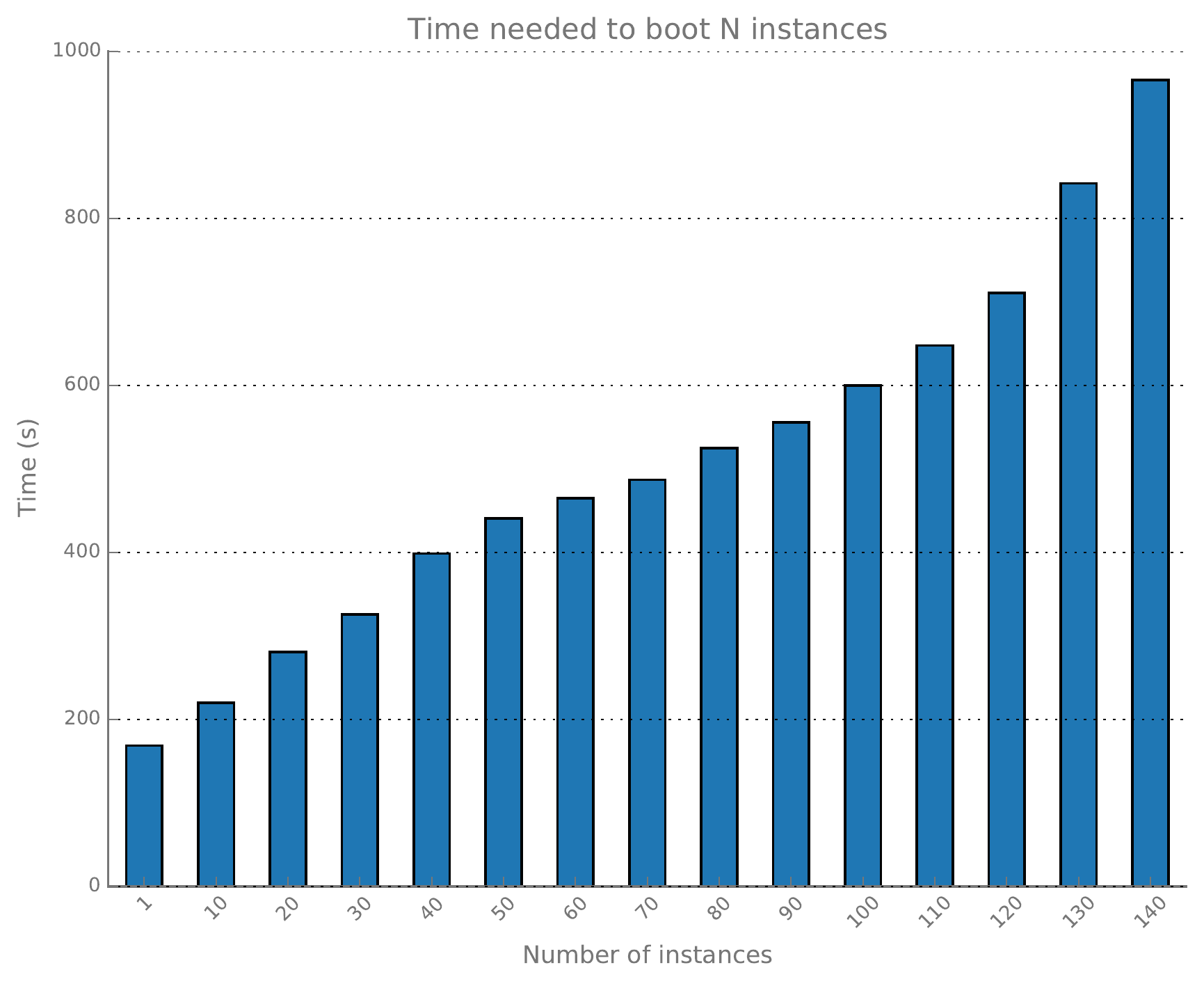}
    \caption{Time needed to boot the number of requested instances. Tests were
    performed in a dedicated infrastructure based on OpenStack with 35 hosts
    over a 1GbE network with an image of 2GB.}
    \label{fig:start time}
\end{figure}

When a request is made, the corresponding images have to be distributed from
the {\color{blue}catalogue} to the compute nodes that will host the virtual
machines.  If the {\color{blue}catalogue} repository is not shared or the image
is not already cached by the compute nodes, this distribution will introduce a
penalty on the start time of the requested nodes. This overhead can be quite
significant {\color{blue}\cite{Ramakrishnan2011}}
and has a large influence in the startup time for a request. This
is specially true when large \cite{Mao2012} requests are made by a user.
Figure~\ref{fig:start time} shows this effect in an OpenStack test
infrastructure. The 2GB images were distributed using HTTP transfers to 35
hosts over a 1GbE network interconnect.  As it can be seen, the time needed to
get all the machines within a single request increased with the size of the
request.

Parallel applications are common in scientific workloads, so a mechanism
should be provided to ensure that these large request are not penalized by this
transfer and deployment time. Users requiring interactivity cannot afford to
wait for several minutes for an instance to be spawned, since interactivity
implies immediateness. This is specially important for the co-allocation of
instances, as described in Section~\ref{sec:challenges:co allocation}, since
the VM provision time may impact in the delivery time to the users, hindering
the co-allocation of resources.

\subsubsection{Short startup overhead open challenges}
\begin{itemize}
    \item How to deploy the images into the nodes in an efficient way.
    \item How to deal with spikes on the requests, so that the systems are not
        saturated transmitting the images into a large number of nodes.
    \item How to implement cache mechanisms in the nodes, implementing sanity
        checks so that similar workloads are not constrained into a few nodes.
    \item How to forecast workloads, so that images can be pre-deployed,
        anticipating the user's requests.
\end{itemize}

\section{Conclusions}
\label{sec:conclusions}

In this paper we have depicted and elaborated on the resource allocation
open challenges for cloud frameworks, based on the analysis of scientific
applications requirements. In this context, we have identify cloud providers
as Science Clouds, since they might not have the same expectations, requirements
and challenges as any other private or commercial cloud infrastructure.

Cloud Management Frameworks {\color{blue}(CMFs)} are normally being developed taking into account
the point of view of a commercial provider, focusing on satisfying the industry
needs, but not really fulfilling academia demands. Scientific workloads are
considered as high-performance computing tasks that need strong requirements.
Some of them were tackled in previous computing paradigms and now there is the
need to address them in Science Clouds. Other resource allocation requirements
identified in this paper are inherent to cloud computing and would provide the
predictability that cloud frameworks currently lack. These requirements
naturally evolve into challenges that, as the time of writing, appear as
obstacles for moving certain scientific workflows to Science Clouds.


The cloud is not a silver bullet for scientific users, but rather a new
paradigm that will enter the ecosystem. In the upcoming years scientific
computing datacenters have to move towards a mixed and combined model, where a
given user will have access to the more \emph{traditional} computational power,
but also they should provide their users with additional cloud power that will
complement the former computing infrastructures. These Science Clouds should
need to be tuned to accommodate the demands of the user communities supported.
This way, either the users the users will benefit from a richer environment,
and resource providers can get a better utilization of their resources, since
they will allow for new execution models that are currently not available.

\bibliography{references}

\end{document}